\newcommand{\nin}{\noindent}

\newcommand{\non}{\nonumber}

\newcommand{\yskip}{\penalty -50 \vskip 3pt plus 3pt minus 4pt}

\newcommand{\qua}{\ {\vbox{\hrule\hbox{\vrule\hbox to 4pt{\vbox to 8pt{}}
\vrule}\hrule}}\ }
\newcommand{\blank}{\rm{\vbox{\hbox{\vrule\hbox to 3pt{\vbox to 2.5pt{}}\vrule}\hrule}}}

\newcommand{\co}[2]{\par\yskip
                    \noindent\hangindent #1cm
                    \hbox to #1cm
                    {#2\hss\hskip 10pt}}
\newcommand{\cond}[1]{\par\yskip
                   \hangindent 0.7cm \noindent
                   \hbox to 0.7cm{#1\hfill}}
\newcommand{\condition}[1]{\par\yskip
                   \hangindent 30pt \noindent
                   \hbox to 30pt{\hss#1\hskip10pt}}
\newcommand{\condvar}[2]{\par\yskip
                   \hangindent #1cm \noindent
                   \hbox to #1cm{\hss#2\hskip10pt}}

\newtheorem{@theorem}{Theorem}
\newcommand{\theorem}[1]{\begin{@theorem}#1\end{@theorem}}

\newtheorem{@satz}{Satz}
\newcommand{\satz}[1]{\begin{@satz}#1\end{@satz}}

\newtheorem{@claim}{Claim}
\newcommand{\claim}[1]{\begin{@claim}#1\end{@claim}}

\newtheorem{@behauptung}{Behauptung}
\newcommand{\behauptung}[1]{\begin{@behauptung}#1\end{@behauptung}}

\newtheorem{@proposition}{Proposition}
\newcommand{\proposition}[1]{\begin{@proposition}#1\end{@proposition}}

\newtheorem{@lemma}{Lemma}
\newcommand{\lemma}[1]{\begin{@lemma}#1\end{@lemma}}

\newtheorem{@corollary}{Corollary}
\newcommand{\corollary}[1]{\begin{@corollary}#1\end{@corollary}}

\newtheorem{@korollar}{Korollar}
\newcommand{\korollar}[1]{\begin{@korollar}#1\end{@korollar}}

\newtheorem{@definition}{Definition}
\newcommand{\definition}[1]{\begin{@definition}#1\end{@definition}}

\newtheorem{@assumption}{Assumption}
\newcommand{\assumption}[1]{\begin{@assumption}#1\end{@assumption}}

\newtheorem{@annahme}{Annahme}
\newcommand{\annahme}[1]{\begin{@annahme}#1\end{@annahme}}

\newtheorem{@remark}{Remark}
\newcommand{\remark}[1]{\begin{@remark}#1\end{@remark}}

\newtheorem{@bemerkung}{Bemerkung}
\newcommand{\bemerkung}[1]{\begin{@bemerkung}#1\end{@bemerkung}}

\newtheorem{@example}{Example}
\newcommand{\example}[1]{\begin{@example}#1\end{@example}}

\newtheorem{@beispiel}{Beispiel}
\newcommand{\beispiel}[1]{\begin{@beispiel}#1\end{@beispiel}}

\def\IIbox#1#2#3 #4 #5{\noindent\hbox{\relax
  \vtop{\hsize #1cm\noindent\strut #4}\hskip #2cm
  \vtop{\hsize #3cm\noindent\strut #5}\hss}\eol}

\newcommand{\gl}{\lower4pt\hbox{${{\displaystyle >}\atop{\displaystyle <}}\atop\/ $}}

\renewcommand{\lg}{\lower4pt\hbox{${{\displaystyle <}\atop{\displaystyle >}}\atop\/ $}}

\newcommand{\sge}{\mathop{\lower 2pt\hbox{$\buildrel >\over =$}}}

\newcommand{\sle}{\mathop{\lower 2pt\hbox{$\buildrel <\over =$}}}

\newcommand{\slg}{\lower1.6truept\hbox{${{{\scriptstyle >}\atop{\scriptstyle =}}\atop {\scriptstyle <}}$}}

\newcommand{\sgl}{\lower1.6truept\hbox{${{{\scriptstyle <}\atop{\scriptstyle =}}\atop {\scriptstyle >}}$}}

\newcommand{\eax}{{{\lower4truept\hbox{$=$}\atop {\textstyle x}}\atop \/}}

\newcommand{\notprecsim}{\hskip 2 pt
                \not\kern -0.3 em{\hbox{$\precsim$}}
                \hskip 2 pt }
\newcommand{\notsuccsim}{\hskip 2 pt
                \not\kern -0.3 em{\hbox{$\succsim$}}
                \hskip 2 pt     }
\newcommand{\precsimi}{\hskip 2 pt
              \hbox{$\precsim_{\kern -0.6 em\raise 0.4 ex\hbox{
              $\scriptstyle i$}}$}
              \hskip 2 pt        }
\newcommand{\succsimi}{\hskip 2 pt
              \hbox{$\succsim_{\kern -0.6 em\raise 0.4 ex\hbox{
              $\scriptstyle i$}}$}
              \hskip 2 pt}
\newcommand{\precsima}{\hskip 2 pt
              \hbox{$\precsim_{\kern -0.6 em\raise 0.1 ex\hbox{
              $\scriptstyle a$}}$}
              \hskip 2 pt}
\newcommand{\precsimA}{\hskip 2 pt
              \hbox{$\precsim_{\kern -0.6 em\raise 0.1 ex\hbox{
              $\scriptstyle A$}}$}
              \hskip 2 pt}
\newcommand{\succsima}{\hskip 2 pt
              \hbox{$\succsim_{\kern -0.6 em\raise 0.1 ex\hbox{
              $\scriptstyle a$}}$}
              \hskip 2 pt}
\newcommand{\succsimA}{\hskip 2 pt
              \hbox{$\succsim_{\kern -0.6 em\raise 0.1 ex\hbox{
              $\scriptstyle A$}}$}
              \hskip 2 pt}
\newcommand{\precsimb}{\hskip 2 pt
              \hbox{$\precsim_{\kern -0.6 em\raise 0.1 ex\hbox{
              $\scriptstyle b$}}$}
              \hskip 2 pt}
\newcommand{\precsimB}{\hskip 2 pt
              \hbox{$\precsim_{\kern -0.6 em\raise 0.1 ex\hbox{
              $\scriptstyle B$}}$}
              \hskip 2 pt}
\newcommand{\succsimb}{\hskip 2 pt
              \hbox{$\succsim_{\kern -0.6 em\raise 0.4 ex\hbox{
              $\scriptstyle b$}}$}
              \hskip 2 pt}
\newcommand{\succsimB}{\hskip 2 pt
              \hbox{$\succsim_{\kern -0.6 em\raise 0.4 ex\hbox{
              $\scriptstyle B$}}$}
              \hskip 2 pt}
\newcommand{\precsimn}{\hskip 2 pt
              \hbox{$\precsim_{\kern -0.6 em\raise 0.1 ex\hbox{
              $\scriptstyle n$}}$}
              \hskip 2 pt }
\newcommand{\succsimn}{\hskip 2 pt
              \hbox{$\succsim_{\kern -0.6 em\raise 0.1 ex\hbox{
              $\scriptstyle n$}}$}
              \hskip 2 pt }
\newcommand{\succsimai}{\hskip 2 pt
               \hbox{$\succsim_{\kern -0.6 em\raise 0.4 ex\hbox{
               $\scriptstyle(a,i)$}}$}
               \hskip 2 pt }

\newcommand{\C}{\hbox{C\hskip-0.5em\lower-0.1ex\hbox{\vrule
                      height1.34ex width0.07em }}\hskip0.50em}

\newcommand{\G}{\hbox{G\hskip-0.525em\lower-0.081ex\hbox{\vrule
                      height1.4ex width0.07em }}\hskip0.50em}

\renewcommand{\O}{\hbox{O\hskip-0.525em\lower-0.095ex\hbox{\vrule
                      height1.45ex width0.07em}}\hskip0.50em}

\newcommand{\Q}{\hbox{Q\hskip-0.525em\lower-0.097ex\hbox{\vrule
                      height1.47ex width0.07em}}\hskip0.50em}

\newcommand{\U}{\hbox{U\hskip-0.45em\lower-0.02ex\hbox{\vrule
                height1.54ex width0.07em}}\hskip0.50em}

\newcommand{\1}{1\hskip-0.28em \text{I}}

\newcommand{\argmin}{\mbox{argmin}}

\documentclass[a4paper,12pt]{article}

\usepackage[]{amstext}
\usepackage[]{amsmath}

\usepackage[]{amssymb}
\usepackage[]{graphicx}
\usepackage[]{threeparttable}
\usepackage[]{lscape}
\usepackage{xcolor}
\usepackage{ulem}
\usepackage{appendix}
\usepackage{biblatex}
\usepackage{ragged2e}

\newcommand*{\QEDB}{\hfill\ensuremath{\square}}%

\begin{document}

\author{Munir Hiabu\footnote{University of Copenhagen, Department of Mathematical Sciences, E--mail: mh@math.ku.dk}\\
Simon M.S. Lo\footnote{The United Arab Emirates University, United Arab Emirates, Department of Innovation in Government and Society, E--mail: losimonms@yahoo.com.hk} \\
Ralf A. Wilke\footnote{Copenhagen Business School, Department of Economics, E--mail: rw.eco@cbs.dk}}
\title{Identifiability and estimation of the competing risks model under exclusion restrictions}
\vspace{1.3cm}
\maketitle
\thispagestyle{empty}

\begin{abstract}
The non-identifiability of the competing risks model requires researchers to work with restrictions on the model to obtain informative results. We present a new identifiability solution based on an exclusion restriction. Many areas of applied research use methods that rely on exclusion restrcitions. It appears natural to also use them for the identifiability of competing risks models. By imposing the exclusion restriction couple with an Archimedean copula, we are able to avoid any parametric restriction on the marginal distributions. We introduce a semiparametric estimation approach for the nonparametric marginals and the parametric copula. Our simulation results demonstrate the usefulness of the suggested model, as the degree of risk dependence can be estimated without parametric restrictions on the marginal distributions.
\end{abstract}

\section{Introduction}
Competing risks duration models are routinely applied in many disciplines, including biostatistics, mechanical engineering, economics and social sciences. While the non-identifiability of the competing risks model (Cox, 1962, Tsiatis, 1975) complicates informative empirical analysis, a series of contributions has obtained identification results under different sets of restrictions (Heckman and Honor\'{e}, 1989;  Aabring and Van den Berg, 2003; Lee, 2006; Lee and Lewbel, 2013; Wang, 2023). This paper contributes to the literature by presenting a new identifiability result for a general class of competing risks models that is obtained under exclusion restrictions. While exclusion restrictions are commonly used in statistical models, we present the first identifiability result for the competing risks duration model that relies on them. Other instrumental variable models for duration analysis are less about identifiability of the marginals or risk dependencies but to tackle endogeneity in covariates (e.g. Beyhum et al., 2022; Martinussen and Vansteelandt, 2020; Richardson et al., 2017; Zheng et al., 2017). Others are restricted to a single risk (Robins and Tsiatis, 1991; Aabring and Van den Berg, 2005; Bijwaard and Ridder, 2005; Bijwaard, 2009).

We consider a competing risks duration model that links the marginal distributions of latent competing durations with the help of a copula (Carri\`{e}re, 1995; Zheng and Klein, 1995). Copula functions are increasingly popular for modelling risk dependence as there is a link to frailty modelling. See Emura et al. (2019) for extensive coverage of advanced statistical models that incorporate heterogeneity of a population by means of frailty and dependence between competing risks in terms of copulas.  Ha et al. (2019) and Lo et al. (2017) establish the link between frailty and the copula. This paper focuses on Archimedean copulas as it simplifies numerical analyses. It demonstrates that the competing risks model is identifiable under exclusion restrictions. A similar observation has been made by Lee and Park (2023) in the context of the extended Roy model. The main advantage of our approach compared to existing identifiability results is that it avoids functional form restrictions on the marginal distributions, maintaining their nonparametric nature, while the degree of risk dependence does not need to be known or assumed. A simulation study provides evidence of the practicality of the suggested approach.

The paper is organised as follows. Section \ref{S:Model} presents the model including the main assumptions. Section \ref{S:Ident} contains the identification result. A nonparametric estimation approach is presented in Section \ref{s:est}. Section \ref{s:sim} presents simulation results which confirm the theoretical findings.

\section{The model \label{S:Model}}
There are two competing risks with corresponding durations $T_{1}$ and $T_{2}$. Observable are the duration to the first failure and the corresponding cause of failure, denoted by $T=\min_j\{T_{j}\}$ and $\Delta = \argmin_j \{T_{j}\}$, $j=1,2$. Let the overall survival function be $\pi(t;z) = \Pr(T_1>t, T_2>t|z)$, where $z\in \mathbb R^d, d\geq 2$ is a vector of observable covariates. We focus here on the case of continuous covariates and we will impose smoothness conditions on $\pi(t;z)$.

\assumption{\label{A:1}(Exclusion restriction)
(i) The marginal survival function for the second risk, $S_2= \Pr( T_2>t|z)$,  does not depend on $z_1$ and the marginal survival function for the first risk, $S_1= \Pr( T_1>t|z)$,  does not depend on $z_2$.
(ii) There is a set of points $\mathcal T\times \mathcal Z \subseteq [0,1]\times \mathbb R^d$  such that the survival probability $\pi(t;z)$ as a function in $z$ is twice continuously differentiable for $(t,z) \in \mathcal T\times \mathcal Z$
with non-vanishing partial derivatives  ${\partial \pi(t;z)}/{\partial z_1}$,  ${\partial \pi(t;z)}/{\partial z_2}$.
}


\assumption{\label{A:2}(Archimedean Copula)
(i) $\pi(t;z) = \phi^{-1}_{\theta}(\phi_{\theta}(S_1(t;z))+ \phi_{\theta}(S_2(t;z)))$, where $\phi_{\theta}$ is a differentiable, strictly decreasing, convex function  that depends on a parameter $\theta$ and maps from  $[0,1]$ to $[0, \infty) $, with  $\phi(1) = 0$ and  convention $\phi^{-1}(u) =0$ for $u\geq \phi(0)$.
 (ii) For any $\theta_1>\theta_2$, $\phi'_{\theta_2}(s)/\phi'_{\theta_1}(s)$ is strictly increasing in $s$.}

From Assumption \ref{A:2}, we have
\begin{eqnarray}
\phi_{\theta}(\pi(t;z)) & = & \phi_{\theta}(S_1(t;z))+ \phi_{\theta}(S_2(t;z)). \label{eq1}
\end{eqnarray}
\nin Assumption \ref{A:2}(ii) is compatible with a range of popular one-parameter copulas, including Clayton, Gumbel, Frank, Joe, and Ali-Mikhail-Haq copulas (Lo and Wilke, 2023, Lemma 3). We focus here on Archimedean copulas as it simplifies the numerical implementation and analysis.

\section{Identifiability under an exclusion restriction \label{S:Ident}}

Taking the partial derivative w.r.t. $z_1$ on both sides of \eqref{eq1} gives
\begin{eqnarray}
\frac{\partial \phi_{\theta}(\pi(t;z))}{\partial \pi} \frac{\partial \pi(t;z)}{\partial z_1} & = & \frac{\partial \phi_{\theta}(S_1(t;z))}{\partial S_1} \frac{\partial S_1(t;z)}{\partial z_1}. \label{eq2}
\end{eqnarray}
Taking the partial derivative of the LHS of (\ref{eq2}) w.r.t. $z_2$, we obtain
\begin{eqnarray}
\frac{\partial^2 \phi_{\theta}(\pi(t;z))}{\partial \pi^2} \frac{\partial \pi(t;z)}{\partial z_1}\frac{\partial \pi(t;z)}{\partial z_2} +  \frac{\partial \phi_{\theta}(\pi(t;z))}{\partial \pi} \frac{\partial^2 \pi(t;z)}{\partial z_1 \partial z_2 }&=& 0. \label{eq3}
\end{eqnarray}
\nin The RHS of (\ref{eq3}) is zero because the RHS of (\ref{eq2}) does not depend on $z_2$ due to Assumption \ref{A:1}(ii). Rearranging (\ref{eq3}) yields
\begin{eqnarray}
\frac{\partial^2 \phi_{\theta}(\pi(t;z))}{\partial \pi^2}  \bigg/ \frac{\partial \phi_{\theta}(\pi(t;z))}{\partial \pi} &=& - \frac{\partial^2 \pi(t;z)}{\partial z_1 \partial z_2 } \bigg/ \bigg(\frac{\partial \pi(t;z)}{\partial z_1}\frac{\partial \pi(t;z)}{\partial z_2}\bigg). \label{eq4}
\end{eqnarray}

\theorem{Under Assumptions \ref{A:1} and \ref{A:2}, the dependence parameter $\theta$ is identifiable.  \label{L:1}}
\nin The proof of Theorem \ref{L:1} corresponds to showing that there is a unique solution for $\theta$ in (\ref{eq4}).
\paragraph{Proof of Theorem \ref{L:1}:}
Assume that  (\ref{eq4}) holds for two different values $\theta_1,\theta_2$. Then
\begin{eqnarray}
 - \frac{\partial^2 \pi(t;z)}{\partial z_1 \partial z_2 } \bigg/ \bigg(\frac{\partial \pi(t;z)}{\partial z_1}\frac{\partial \pi(t;z)}{\partial
 z_2}\bigg)&=& \frac{\partial^2 \phi_{\theta_1}(\pi(t;z))}{\partial \pi^2}  \bigg/ \frac{\partial \phi_{\theta_1}(\pi(t;z))}{\partial \pi}  \notag  \\ &=&  \frac{\partial^2 \phi_{\theta_2}(\pi(t;z))}{\partial \pi^2}  \bigg/ \frac{\partial \phi_{\theta_2}(\pi(t;z))}{\partial \pi}. \label{L1p}
\end{eqnarray}
We will show that  \eqref{L1p} cannot hold for $\theta_1 > \theta_2$. Then by symmetry also $\theta_1 < \theta_2$ cannot hold,
such that  $\theta_1 = \theta_2$.
Define
\begin{eqnarray*}
\psi(\pi) &:=&  \frac{\partial \phi_{\theta_2}(\pi(t;z))}{\partial \pi} \bigg/ \frac{\partial \phi_{\theta_1}(\pi(t;z))}{\partial \pi}. 
\end{eqnarray*}
\nin Taking the derivative 
gives
\begin{eqnarray*}
\psi'(\pi) &=&
\frac{\frac{\partial \phi_{\theta_1}(\pi(t;z))}{\partial \pi} \frac{\partial^2 \phi_{\theta_2}(\pi(t;z))}{\partial \pi^2} - \frac{\partial \phi_{\theta_2}(\pi(t;z))}{\partial \pi} \frac{\partial^2 \phi_{\theta_1}(\pi(t;z))}{\partial \pi^2}}{\left[\frac{\partial \phi_{\theta_1}(\pi(t;z))}{\partial \pi}\right]^2}.
\end{eqnarray*}
\nin Assumption \ref{A:2}(ii) states that $\psi'(\pi)>0$, hence the numerator 
 must be greater than zero, implying
\begin{eqnarray}
 \frac{\partial^2 \phi_{\theta_2}(\pi(t;z))}{\partial \pi^2}  \bigg/ \frac{\partial \phi_{\theta_2}(\pi(t;z))}{\partial \pi}  &>&
\frac{\partial^2 \phi_{\theta_1}(\pi(t;z))}{\partial \pi^2}  \bigg/ \frac{\partial \phi_{\theta_1}(\pi(t;z))}{\partial \pi}.
\non
\end{eqnarray}
Hence, \eqref{L1p} cannot hold for $\theta_1 > \theta_2$.
\QEDB \\

In the following we present the identifying conditions for several popular copulas.

\example{(Clayton copula): $\phi_{\theta}(s) = (s^{-\theta}-1)/\theta$ with $\theta \in[-1,\infty)\backslash \{0 \}$. For $\theta=0$, it is the independence copula with $\phi_{0}(s) = \log(s)$. Equation (\ref{eq4}) becomes
\begin{eqnarray}
(\theta+1)/\pi(t;z) = \frac{\partial^2 \pi(t;z)}{\partial z_j \partial z_l } \bigg/ \bigg(\frac{\partial \pi(t;z)}{\partial z_j}\frac{\partial \pi(t;z)}{\partial z_l}\bigg)
\end{eqnarray}
or
\begin{eqnarray}
\theta &=& \pi(t;z)\frac{\partial^2 \pi(t;z)}{\partial z_j \partial z_l } \bigg/ \bigg(\frac{\partial \pi(t;z)}{\partial z_j}\frac{\partial \pi(t;z)}{\partial z_l}\bigg)-1
\end{eqnarray}
for all $(t,z) \in \mathcal T\times \mathcal Z$. The parameter $\theta$ is therefore locally identified. In an application with unknown $\pi(t;z)$ one can use a nonparametric model for estimation. Estimation of $\theta$ can be through averaging, exploiting the sample variation in $t$ and $z$:
\begin{eqnarray}
\theta &=& E_{T,Z}\left[\pi(t;z)\frac{\partial^2 \pi(t;z)}{\partial z_j \partial z_l } \bigg/ \bigg(\frac{\partial \pi(t;z)}{\partial z_j}\frac{\partial \pi(t;z)}{\partial z_l}\bigg)\right]-1. \label{theta}
\end{eqnarray}}

\example{(Gumbel copula): $\phi_{\theta}(s) = (-\log s)^{\theta}$ with  $\theta \in(1,\infty)$. We have
\begin{eqnarray}
\frac{1}{\pi(t;z)}\bigg(\frac{\theta-1}{\log \pi(t;z)}-1\bigg) = \frac{\partial^2 \pi(t;z)}{\partial z_1 \partial z_2 } \bigg/ \bigg(\frac{\partial \pi(t;z)}{\partial z_1}\frac{\partial \pi(t;z)}{\partial z_2}\bigg),
\end{eqnarray}
\nin for all $(t,z) \in \mathcal T\times \mathcal Z$. Since the LHS is strictly decreasing in $\theta$, there is a unique solution
\begin{eqnarray*}
\theta = ( A(t;z) +1) B(t;z) +1
\end{eqnarray*}
with
$$A(t;z) := \pi(t;z)  \frac{\partial^2 \pi(t;z)}{\partial z_1 \partial z_2 }  \bigg/\bigg(\frac{\partial \pi(t;z)}{\partial z_1} \log \frac{\partial \pi(t;z)}{\partial z_2}\bigg)$$ and
$$B(t;z) := -\log \pi(t;z).$$
Estimation of $\theta$ can be once again done by averaging
\begin{eqnarray}
\theta = E_{T,Z}\left[( A(t;z) +1) B(t;z)\right] +1.
\end{eqnarray}}

\example{(Frank copula): $\phi_{\theta}(s) = -\log \Big(\frac{\exp(-\theta t)-1}{\exp(-\theta)-1}\Big)$ with $\theta \in[-\infty,\infty)\backslash \{0 \}$. We have
\begin{eqnarray}
\frac{\theta}{\exp(-\theta \pi(t;z))-1} = \frac{\partial^2 \pi(t;z)}{\partial z_1 \partial z_2 } \bigg/ \bigg(\frac{\partial \pi(t;z)}{\partial z_1}\frac{\partial \pi(t;z)}{\partial z_2}\bigg),
\end{eqnarray}
\nin  for all $(t,z) \in \mathcal T\times \mathcal Z$. Since the LHS is strictly increasing in $\theta$, there is a unique, although non-analytical, solution.}

\section{Estimation \label{s:est}}
The previous section has presented identifiability results for different Archimedean copulas. They all have in common that no parametric assumptions on $S_1(t;z)$ and $S_2(t;z)$ have been made. In practice, the functional form of $\pi(t;z)$ is unknown. For estimation in a regression setting, it is natural to assume a (semi-)parametric model for the cause specific hazards (CSHs) to impose some structure on $\phi_{\theta}$ for higher dimensional settings. 
In our case it is however not clear how a structural assumption impacts the copula assumption and what set of functions would satisfy both.
Unfortunately, the analytical link between restrictions on $\pi(t;z)$ and $\phi_{\theta}$ is difficult. To avoid any kind of misspecification, we take a nonparametric route in the following. The estimation is by means of sample analogues of the copula specific  solutions for $\theta$ as given in Section \ref{S:Ident}. These require nonparametric estimates of $\pi(t;z)$, ${\partial \pi(t;z)}/{\partial z_j}$ for $j\in\{1,2\}$ and $\partial^2 \pi(t;z)/\partial z_1 \partial z_2$.

Nonparametric estimation of survival models in presence of continuous covariates has been considered in the literature (Beran, 1981; Dabrowska, 1987; Wichert and Wilke, 2008, Selingrov\'{a} et al., 2014), although we are not aware of an estimator for $\pi(t;z)$ and its partial and cross derivatives w.r.t. $z$. These are therefore suggested in the following.

Suppose $(t_i,z_i)$ for $i\in\{1,\ldots,N\}$ is a random sample. The starting point is a Kernel estimator for $\pi((t;z))= \Pr(T>t|z)$ which smoothes in $z$. First, we remark $\pi(t;z)=\Pr(T>t,z)/\Pr(z)$. These probabilities are estimated by
\begin{eqnarray*}
\hat{\pi}(t;z) &=& \frac{\sum_{i=1}^N \1\{T_i> t\} K_{\mathbf{h}}(z-z_i)}{\sum_{i=1}^N K_{\mathbf{h}}(z-z_i)},
\end{eqnarray*}
\noindent where $K_{h_k}(z_k-z_{ki})=h^{-1}K(h^{-1}(z_k-z_{ki}))$ for some bandwidth $h_k>0$ and  a Kernel function $K$. Furthermore, $K_{\mathbf{h}}(z-z_i)=\prod_{k=1}^d K_{h_k}(z_k-z_{ki})$ is the $d$ dimensional product kernel with $\mathbf{h}=(h_1,\ldots,h_d)$. The estimator can be easily extended to accommodate situations with independent right censoring due to the end of the observation period or due to random dropouts.

For the estimator of the partial derivatives of $\pi(t;z)$, we suggest taking the partial derivatives of $\hat{\pi}(t;z)$ w.r.t. $z_k$:
\begin{eqnarray*}
\hat{\pi}_{z_k}(t;z) &=& \frac{\sum_{i=1}^N \1\{T_i> t\} K'_{\mathbf{h},k}(z_k-z_{ki})\prod_{j\neq k} K_{h_j}(z_j-z_{ji})\sum_{i=1}^N K_{\mathbf{h}}(z-z_i)}{\left(\sum_{i=1}^N K_{\mathbf{h}}(z-z_i)\right)^2} \\
    && -  \frac{\sum_{i=1}^N K'_{\mathbf{h},k}(z_k-z_{ki})\prod_{j\neq k} K_{h_j}(z_j-z_{ji})\sum_{i=1}^N \1\{T_i> t\} K_{\mathbf{h}}(z-z_i)}{\left(\sum_{i=1}^N K_{\mathbf{h}}(z-z_i)\right)^2},
\end{eqnarray*}
\noindent where $K'_{\mathbf{h},k}$ and $K'_{h_k}$ are the first derivative of $K_{\mathbf h}$ with respect to $z_k$. Another route could be to do the equivalent of average derivative estimation as in H\"ardle and Stoker (1989).

The estimator for the cross derivative is more complicated but can be obtained in straight forward manner. For this, we define:
\begin{eqnarray*}
a(t,z)&:=& \sum_{i=1}^N \1\{T_i> t\} K_{\mathbf{h}}(z-z_i),\\
b(z)&:=& \sum_{i=1}^N K_{\mathbf{h}}(z-z_i),
\end{eqnarray*}
with $\partial a(t;z)/\partial z_k= \sum_{i=1}^N \1\{T_i> t\} K'_{\mathbf{h},k}(z-z_i)$ and $\partial b(z)/\partial z_k=\sum_{i=1}^N K'_{\mathbf{h},k}(z-z_i)$. The cross derivatives are rather simple because we use the product kernel as the product of univariate kernel functions such that:
\begin{eqnarray*}
  \frac{\partial^2 a(t,z)}{\partial z_k \partial z_l}  &=& \sum_{i=1}^N \1\{T_i> t\} K'_{h_k}(z_k-z_{ki})K'_{h_l}(z_l-z_{li})\prod_{j\neq k,l} K_{h_j}(z_j-z_{ji})   \\
  \frac{\partial^2 b(z)}{\partial z_k \partial z_l}  &=&  \sum_{i=1}^N K'_{h_k}(z_k-z_{ki})K'_{h_l}(z_l-z_{li})\prod_{j\neq k,l} K_{h_j}(z_j-z_{ji}),
\end{eqnarray*}
where $K'_{h_k}$ is the first derivative of $K_{h_k}$ with respect to $z_k$. These quantities are used to compute
\begin{eqnarray*}
\hat{\pi}_{z_k,z_l}(t;z) &=& \frac{\frac{\partial^2 a(t,z)}{\partial z_k \partial z_l}}{b(z)}-\frac{\frac{\partial b(z)}{\partial z_k}\frac{\partial a(t;z)}{\partial z_l}+\frac{\partial a(t;z)}{\partial z_k}\frac{\partial b(z)}{\partial z_l}+\frac{\partial^2 b(z)}{\partial z_k \partial z_l}a(t,z)}{b(z)^2}
 +\frac{2\frac{\partial b(z)}{\partial z_k} a(t,z)\frac{\partial b(z)}{\partial z_l}}{b(z)^3}
\end{eqnarray*}

Let $\hat{\theta}$ be the plug-in estimator of $\theta$ by plugging-in the nonparametric estimators into the closed form of $\theta$.  In the following, we show that the suggested estimators are consistent under the following assumptions.
\begin{enumerate}
\item[A1] The kernel $K$ is symmetric, has bounded support and has finite second moment.
\item[A2] The density function of $(T,Z)$, $p^{T,Z}$, is strictly positive and four times continuously differentiable.
\item[A3] The bandwidth $h=h(n)$ satisfies $h\rightarrow 0$ and  ${nh^{2+d}}\rightarrow \infty$.
\end{enumerate}

\begin{corollary}{\label{cor1}} Assume A1--A3. The plug-in estimators of $\theta$ in Examples 1--3 of Section \ref{S:Ident} are weakly consistent.
\end{corollary}
\paragraph{Proof of Corollary \ref{cor1}:}
From Lemma \ref{lem1} in the Appendix, we conclude that $\hat{\pi}(t;z)$, $\hat{\pi}_{z_k}(t;z)$ for $k\in \{1,2\}$, and $\hat{\pi}_{z_k,z_l}(t;z)$ for $k\neq l$ are weakly consistent estimators.  The claim follows by applying the continuous mapping theorem. \QEDB

\section{Simulations \label{s:sim}}
We simulate data for a known model and compare the estimation results with their true values to assess the finite sample performance. The covariates $z=(z_1, z_2)'$ are $z_j \sim N(0,0.5)$ for $j\in \{1,2\}$. We use a Weibull model with $\Lambda_j(t)=\lambda_j t^{\eta_j}$ with $(\lambda_1, \eta_1, \lambda_2, \eta_2) = (0.5, 1, 1, 1)$. Notice that $\lambda_j(t)=\lambda_j \eta_j t^{\eta_j-1}$ and $h_j(t,\eta_j)=\eta_j/t$ in this model. As covariate function we use $\exp(z_j\beta_{z_j})$ with $(\beta_{z_1}, \beta_{z_2})=(1, 1)$. Therefore $S_j(t;z_j)=\exp(-\Lambda_j(t)\exp(z_j\beta_{z_j}))$ for $j=1,2$. $C(s_1,s_2;\theta)$ is the Clayton copula with Kendall's $\tau=0.2$ $(\theta=0.5)$ and $\theta=2\tau/(1-\tau)$.

In this model, $\pi(t;z)$ and its derivatives are given by:
\begin{eqnarray*}
\pi(t;z) & = & (S_1^{-\theta}+S_2^{-\theta}-1)^{-1/\theta}\\
\frac{\partial \pi(t;z)}{\partial z_j} &=& S_j^{-(\theta+1)}(S_1^{-\theta}+S_2^{-\theta}-1)^{-(1+1/\theta)}\times \frac{\partial S_j}{\partial z_j} \\
\frac{\partial^2 \pi(t;z)}{\partial z_j\partial z_l} &=& (1+\theta)(S_1^{-\theta}+S_2^{-\theta}-1)^{-(2+1/\theta)}S_j^{-(\theta+1)}S_l^{-(\theta+1)}\times \frac{\partial S_j}{\partial z_j} \frac{\partial S_j}{\partial z_l}
\end{eqnarray*}
with simplified notation $S_j(t,z_j)=S_j$ and
\begin{eqnarray*}
\frac{\partial S_j}{\partial z_j}&=&  -S_j(t,z_j)\Lambda_j(t)\exp(z_j\beta_{z_j})\beta_{z_j}.
\end{eqnarray*}

It is remarked: $C(s_1,s_2;\theta)=(s_1^{-\theta}+s_2^{-\theta}-1)^{-1/\theta}$. The conditional copula is
$C(s_2;s_1)=\partial C /\partial s_1$ and $s_2=C^{-1}(v_2;s_1)$.

The data are simulated with the following procedure:

\begin{enumerate}
\item Generate two uniform random variables $s_1$ and $v_2$ on $[0,1]$ with $N$ independent random draws.

\item Obtain $N$ realisations of $s_2$ by
\[s_2=\left(1-s_1^{-\theta}+(v_2s_1^{\theta+1})^{-\theta/(\theta+1)}\right)^{-1/\theta}.\]

\item Generate $z_j$ and obtain durations $t_j$ given $z_j$ for $j=1,2$ by inverting the marginal Weibull survival:
\[t_j = \left(-\log(s_j)/\lambda_j/\exp(z_j\beta_{z_j})\right)^{1/\eta_j}.\]

\item Generate observed minimum duration and observed risk by relating $t_1$ and $t_2$.

\end{enumerate}

\begin{figure}[!htbp]
    \centering
    \includegraphics[scale=1.00]{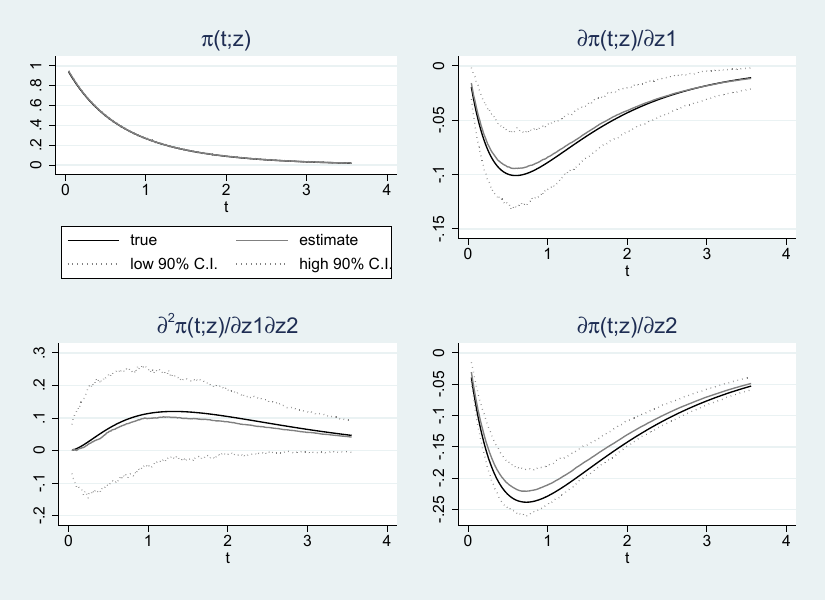}
	\caption{Simulation results: Mean of nonparametric estimates of $\pi(t;\bar{z})$, $\frac{\partial \pi(t;\bar{z})}{\partial z_j}$ for $j\in\{1,2\}$ and $\partial^2 \pi(t;\bar{z})/\partial z_1 \partial z_2$ as functions of $t$ with true functions and confidence bands.}
	\label{F:res}
\end{figure}

We generate $500$ random samples of size $100,000$. We apply the nonparametric estimation procedure of Section \ref{s:est}, where we choose $h_1=h_2=0.3$ and use the Epanechnikov Kernel. To speed up the process, estimation is done on a fixed grid for $T$ with 500 grid points $\tilde{t}_s$ and conditional to the sample average of $z_j$ ($\bar{z}_j$) for $j\in \{1,2\}$. The estimated $\pi(t;\bar{z})$, $\partial \pi(t;\bar{z})/\partial z_j$ for $j\in\{1,2\}$ and $\partial^2 \pi(t;\bar{z})/\partial z_1 \partial z_2$ as functions of $t$ along with confidence bands are displayed in Figure \ref{F:res}. It is evident that there is no systematic bias. It can be also seen that the mean estimate of the partial derivatives, but in particular of the cross derivative have rather wide confidence bands. This suggests that Kernel estimation of derivatives requires large data sets.

\begin{figure}[!htbp]
    \centering
    \includegraphics[scale=0.55]{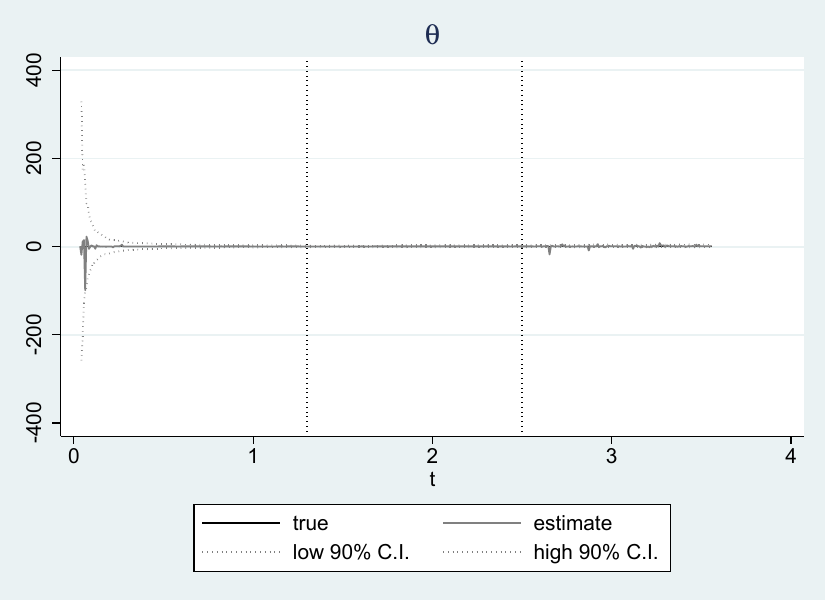}\hspace{1cm}
    \includegraphics[scale=0.55]{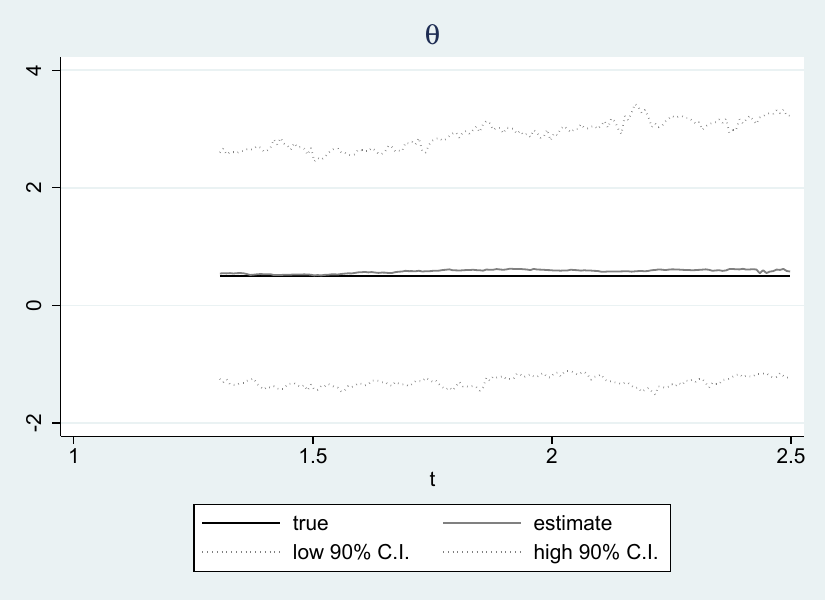}
	\caption{Simulation results: Mean of estimate of $\theta$ as function of $t$ with true value and confidence bands; all grid points (left); trimmed (right). Note: vertical lines in the left panel show the trimming thresholds.}
	\label{F:res2}
\end{figure}

We estimate $\theta$ by a related sample analogue of (\ref{theta}), where we only exploit variation across grid points. Moreover, we allow for trimming the left and right tail of the observed support of $T$:
\begin{eqnarray*}
\hat{\theta} &=&  \left(\sum_{s=1}^{500}\1(\underline{t}\leq\tilde{t}_s\leq \overline{t})\right)^{-1}
\sum_{s=1}^{500} \1(\underline{t}\leq\tilde{t}_s\leq \overline{t}) \left[\hat{\pi}(\tilde{t}_s;\bar{z})\frac{\partial^2 \hat{\pi}(\tilde{t}_s;\bar{z})}{\partial z_j \partial z_l } \bigg/ \bigg(\frac{\partial \hat{\pi}(\tilde{t}_s;\bar{z})}{\partial z_j}\frac{\partial \hat{\pi}(\tilde{t}_s;\bar{z})}{\partial z_l}\bigg)\right]-1\\
&=&\left(\sum_{s=1}^{500}\1(\underline{t}\leq\tilde{t}_s\leq \overline{t})\right)^{-1}
\sum_{s=1}^{500} \1(\underline{t}\leq\tilde{t}_s\leq \overline{t}) \hat{\theta}(\tilde{t}_s).\\
\end{eqnarray*}
We report two sets of results: One for the average over all grid point (no trimming) and one with $\underline{t}=1.3$ and $\overline{t}=2.5$. The trimmed version of the estimator is the average over 170 grid points. We have also tried other averaging, such as taking the average in $t_i$ over all observations instead of the 500 grid points. Because this did not yield an improvement of the results but substantially slowed down the estimation, we decided to use the averaging over grid points and report these results. The trimming limits are chosen by looking at the non-trimmed results for $\hat{\theta}(t)$ as shown in Figure \ref{F:res2} (left). It is evident that estimates become less and less stable for duration approaching zero. A weaker but still visible pattern can be seen for large $t$. In an application, the trimming bounds can be determined on the grounds of a resampling distribution. The distribution of the estimated $\theta$ as a function of $t$ after trimming is shown in Figure \ref{F:res2} (right). It is apparent that the estimate is unbiased for all $t$ but confidence bands are wide. The averaging over the grid points therefore aims at increasing the stability of the estimate.

We report the mean of the 500 $\hat{\theta}$ and 5th and 95th percentiles of the distribution of $\hat{\theta}$ in Table \ref{T:res}. It is apparent that variation in the estimated $\theta$ shrinks drastically compared to Figure \ref{F:res2}. There is a small smoothing related bias for the trimmed estimate, though, which is practically not relevant. This can be seen by the conversion of the 5th and 95th percentile of $\hat{\theta}$ into a range of Kendall's-$\tau$ by using $\tau=\theta/(\theta+2)$ for the Clayton copula. This gives $\hat{\tau}\in [0.2064,2363]$, where the true $\tau$ is $0.2$. Our numerical example therefore illustrates that the suggested estimator gives a rather precise estimate of the degree of dependence. Although it also demonstrates that the two dimensional smoothing in the context of derivative estimation demands large sample sizes. A reduction to fewer observations leads to very noisy estimates or systematic smoothing related biases of the derivative estimates and therefore $\hat{\theta}$.

Future work may develop alternative nonparametric estimators for the derivative estimation that give more precise estimates. We extensively explored already in the direction of taking numerical derivatives of $\hat{\pi}$, the use of random forest type algorithms to fit $\hat{\pi}$ or by imposing parametric constraints on $\pi$ such as a semiparametric proportional cause specific hazards model. While the first did not give better numerical results, the second resulted in larger biases for $\hat{\pi}$ and therefore all subsequent estimates. The imposition of constraints on $\pi$ induces restrictions the copula structure, which could be hard to justify in an application.

\begin{table}
\centering
\caption{Simulation results for $\hat{\theta}$.}
	\begin{tabular}{rcc}
		\hline\hline
& no trimming & trimming  \\
\hline
$\hat{\theta}$ &  $0.4688$ & $.5789$       \\
5th, 95th percentile of $\hat{\theta}$ & $[0.1148,1.7768]$ & $[.5202,.6187]$ \\
\hline
\hline
\multicolumn{3}{l}{{\footnotesize Note: True $\theta=0.5$.}}
\end{tabular}
\label{T:res}
\end{table}

{}

\section*{Appendix}

In the following, we will write subscripts $z_l, z_k$ to denote partial derivatives  and $p^Z$ denotes the marginal density of $Z$.
\lemma{
\label{lem1}
Assume A1--A3, it holds in probability
\begin{align*}
 a(t,z) &\rightarrow \pi(t;z)p^Z(z) ,\\
\hat p (z)  &\rightarrow p^Z(z), \\
a_{z_k}(t,z) &\rightarrow  \pi_{z_k}(t;z)p^Z(z) + \pi(t;z)p^{Z}_{z_k}(z),\\
\hat p^{Z}_{z_k}  &\rightarrow p^{Z}_{z_k}, \\
a_{z_k,z_l}(t,z) &\rightarrow \pi_{z_k,z_l}(t;z)p^Z(z) + \pi_{z_k}(t;z)p^{Z}_{z_l}(z) + \pi_{z_l}(t;z)p^{Z}_{z_k}(z) + \pi(t;z)p_{z_k,z_l}^{Z}(z), \\
\hat p^{Z}_{z_k,z_l}  &\rightarrow p^{Z}_{z_k,z_l}.
\end{align*}}

\paragraph{Proof of Lemma \ref{lem1}:}
We only show the convergence of  $ a^{ }_{z_k}(t,z)$. The other cases follow by similar arguments.
We first calculate the expectation.
By applying a change of variables $\{u_j \rightarrow z_j- hu_j\}$ and thereafter a second order Taylor expansion of $p^{T,Z}$ in $u_j,  j\neq k$ we get
\begin{align*}
\mathbb E_{T,Z}[ a^{ }_{z_k}(t,z)]&= \int \1\{s> t\} K'_{\mathbf{h},k}(z_k-u_k)\prod_{j\neq k} K_{h_j}(z_j-u_j)p^{T,Z}(s,u) \mathrm du \mathrm ds \\
&= \int \1\{s> t\} K'_{\mathbf{h},k}(z_k-u_k) p^{T,Z}(s,u) \mathrm du_{k} \mathrm ds + O(h^2)
\end{align*}
Next, via integration by parts
\begin{align*}
\int  & \1\{s> t\} K'_{\mathbf{h},k}(z_k-u_k) p^{T,Z}(s,u) \mathrm du_{k} \mathrm ds\\
&= \int \1\{s> t\} K_{\mathbf{h}} (z_k-u_k) p^{T,Z}_{z_k} (s,z_1,\dots,z_{k-1},z_{k+1}, \dots z_d)\mathrm du_{k} \mathrm ds.
\end{align*}
A final change of variables with a second order Taylor expansion of $p^{T,Z}_{z_k}$ yields
\begin{align*}
\int  & \1\{s> t\} K_{\mathbf{h}}(z_k-u_k) p^{T,Z}_{z_k}(s,z_1,\dots,z_{k-1},z_{k+1}, \dots z_d) \mathrm du_{k} \mathrm ds\\
&= \int \1\{s> t\}  p^{T,Z}_{z_k} (s,z) \mathrm ds + o_p(h^2)=  \pi_{z_k}(t;z)p^Z(z) + \pi(t;z)p^{Z}_{z_k}(z)+ O(h^2).
\end{align*}
For the variance, we have
\begin{align*}
\mathbb V_{T,Z}[\hat a^{ }_{z_k,z_l}(t,z)]&= \frac 1 n \mathbb V_{T,Z}\left [\1\{T_i> t\} K'_{\mathbf{h},k}(z_k-z_{ki})\prod_{j\neq k} K_{h_j}(z_j-z_{ji})\right]\\
&=  \frac 1 n \mathbb E_{T,Z}\left [\left\{\1\{T_i> t\} K'_{\mathbf{h},k}(z_k-z_{ki})\prod_{j\neq k} K_{h_j}(z_j-z_{ji})\right\}^2\right]
+ O(n^{-1})\\
&=  \frac 1 n \int \1\{s> t\} \left\{ K'_{\mathbf{h},k}(z_k-u_k)\prod_{j\neq k} K_{h_j}(z_j-u_j)\right \}^2 p^{T,Z}(s,u) \mathrm du \mathrm ds + O(n^{-1}).
\end{align*}
Applying a change of variables $\{u \rightarrow z hu\}$ and thereafter a Taylor expansion of $p^{T,Z}$ in $u$, we get
\begin{align*}
\mathbb V_{T,Z}[a^{ }_{z_k,z_l}(t,z)]&= \frac 1 n \mathbb V_{T,Z}\left [\1\{T_i> t\} K'_{\mathbf{h},k}(z_k-z_{ki})\prod_{j\neq k} K_{h_j}(z_j-z_{ji})\right]\\
&=  \frac 1 n \mathbb E_{T,Z}\left [\left\{\1\{T_i> t\} K'_{\mathbf{h},k}(z_k-z_{ki})\prod_{j\neq k} K_{h_j}(z_j-z_{ji})\right\}^2\right]
+ O(n^{-1})\\
&=  \frac 1 {nh^{1+d}} \int \1\{s> t\} \left\{ K'_{k}(u_k)\prod_{j\neq k} K(u_j)\right \}^2 p^{T,Z}(s,z) \mathrm du \mathrm ds + O(n^{-1})\\
&=O(n^{-1}h^{-1-d}).
\end{align*}
\QEDB

\end{document}